\documentclass[12pt]{article}
\usepackage{amssymb,amsmath,graphicx,latexsym}\pdfoutput=1
\begin{document}
\title{\textbf{A pseudoscalar at the LHC:\\technicolor vs a fourth family}} 
\author{B.~Holdom\thanks{bob.holdom@utoronto.ca}\\
\emph{\small Department of Physics, University of Toronto}\\[-1ex]
\emph{\small Toronto ON Canada M5S1A7}}
\date{}
\maketitle
\begin{abstract}
Condensates of colored fermions driving electroweak symmetry breaking could give rise to a light pseudoscalar that would show up in the Higgs search at the LHC. The branching ratios of its decay will distinguish it from the Higgs and illuminate its origin. In particular the colored fermions may be the techniquarks of a one family technicolor theory or the quarks of a strongly interacting fourth family. The former has more difficulties in this context.
\end{abstract}

Electroweak symmetry breaking is intrinsically natural if it is due to the condensation of strongly interacting fermions. It is also noteworthy that the condensing fermions need not be exotic fermions; a family of fermions with standard model quantum numbers is quite adequate. One family technicolor models were extensively studied very early \cite{a1,a2}. Although the interest in such models have ebbed and waned over the years, they have not been conclusively eliminated by experimental data. Electroweak precision constraints along with recent constraints from the LHC \cite{a3} point towards the minimal number of technicolors, $N_{TC}=2$. An even more minimal model of electroweak symmetry breaking occurs when the strong gauge interaction responsible for the condensation is itself broken. Then it is possible that the condensing fermions are no more than a standard (although strongly interacting) fourth family \cite{a7,a8,a9,a10}.

The one family technicolor model and the strong fourth family model are similar in many respects. They both predict that the lightest states of the new strong sector are neutral color singlet pseudoscalar mesons that arise naturally as pseudo-Goldstone bosons of the symmetry breaking. The masses of such states are difficult to predict. Meanwhile a light scalar meson with the properties of the Higgs particle is not expected in such theories in the absence of fine tuning.\footnote{There continues to be controversy about whether a light dilaton exists in walking technicolor theories, but this author believes not \cite{a17}.} Since the condensing (techni)quarks carry color as well as charge, the pseudoscalar will acquire a coupling to two gluons as well as two photons \cite{a2}. Thus it can be produced at the LHC and decay to $\gamma\gamma$ in a manner similar to the Higgs. But unlike the Higgs, the pseudoscalar does not have tree level couplings to $WW$ and $ZZ$ and so those branching ratios will be much suppressed. Due to such differences \cite{a14, a15} the LHC is on the verge of settling the question of whether nature adopts a mechanism for electroweak symmetry breaking that is apparently natural and economical. We shall discuss here how the properties of the pseudoscalar state can differentiate between the technicolor and fourth family options.

Let $\hat{\Sigma}$ be the dynamical fermion mass matrix, a matrix in the space of flavors with respect to the strong interaction. Let $\Pi_a(x)$ denote the Goldstone and pseudo-Goldstone bosons (PGBs) that will elevate this matrix to a slowly varying function of spacetime,
\begin{equation}
\Sigma(x)=U(x)\hat{\Sigma}U(x)^T,\quad U(x)=\exp\sum_a\frac{i\Pi_a(x)X_a}{F_a}
.\label{e3}\end{equation}
The number of different $F_a$'s is determined by the number of different representations by which the PGBs transform under the unbroken subgroup. The physical PGBs may correspond to combinations of the $\Pi_a$'s. In particular a physical combination needs to be orthogonal to any Goldstone boson absorbed by a gauge boson. And among the remaining neutral PGBs there may need to be a transformation to the mass eigenstate basis.  If the combination $\sum_a p_a\Pi_a$ is such a physical PGB, denoted by $\Pi$, then the current to which it couples is determined by the matrix $\sum_a p_aX_a/F_a\equiv X/v$ with $v=246$ GeV. This definition of $X$ determines our convention for its normalization.

In both the case of technicolor and the fourth family there is one neutral color singlet Goldstone boson absorbed by a weak gauge boson. The nonabelian technicolor gauge interaction removes a further potential PGB through instanton effects, and this imposes a traceless condition on the PGB generators. For the fourth family we assume no such nonabelian gauge interaction and thus no traceless condition. We also assume that the fourth family quark condensates do not contribute to the mass of any other gauge bosons other than the electroweak gauge bosons. For example the strong interaction could involve a broken U(1) with vector couplings to the fourth family quarks, as discussed in \cite{a11}.

The PGB $\Pi$ of interest shall be required to have a significant coupling to $gg$ through the axial-vector anomaly. For technicolor we shall consider three examples for the content of the mass eigenstate $\Pi$.
\begin{eqnarray}
A&&\overline{Q}\gamma_5Q-3\overline{L}\gamma_5L\\
B&&\overline{D}\gamma_5D-3\overline{E}\gamma_5E\\
C&&\overline{U}\gamma_5U-3\overline{N}\gamma_5N\label{e5}
\end{eqnarray}
Case A was treated in the original one family technicolor model \cite{a1}. Reference \cite{a5} argued that the lightest mass eigenstate would more likely correspond to B (a case also studied in \cite{a4}). We have included case C for completeness.

For the fourth family we shall be interested in the PGB with content closest to $\overline{t'}\gamma_5t'+\overline{b'}\gamma_5b'$. An isotriplet $\overline{t'}\gamma_5t'-\overline{b'}\gamma_5b'$ component will force the state to also have a lepton component to remain orthogonal to the absorbed Goldstone boson. There can also be mixing with an isosinglet leptonic component. For now we shall take $\Pi$ to be purely $\overline{t'}\gamma_5t'+\overline{b'}\gamma_5b'$, and we shall consider deviations from this choice later. The other PGB(s) will have mostly leptonic components, especially since the fourth family lepton masses are likely smaller than the fourth family quark masses. The $\Pi$ properties are influenced by how much the leptons contribute to electroweak symmetry breaking. We consider two limiting cases, with the lepton masses vanishing (fourth family A) and when they are the same as the quark masses (fourth family B). The physical case lies between these two limits.

From these considerations it is sufficient to consider the case that the unbroken subgroup is maximal, in which case all the $F_a$'s are the same. We shall be interested in the case of a symmetric $\hat{\Sigma}$, and for illustration we show the case for one doublet $(\psi_1,\psi_2)$,
\begin{equation}
\hat{\Sigma}=\left.\begin{array}{c}\psi_{1L} \\\psi_{2L} \\\psi^c_{1L} \\\psi^c_{2L}\end{array}\right.\!\!\!\!\left(\begin{array}{cccc}0 & 0 & 1 & 0 \\0 & 0 & 0 & 1 \\1 & 0 & 0 & 0 \\0 & 1 & 0 & 0\end{array}\right)
.\end{equation}
The $X$ generator in this example takes the form 
\begin{equation}
X=\left(\begin{array}{cccc}x_1 & 0 & 0 & 0 \\0 & x_2 & 0 & 0 \\0 & 0 & x_1 & 0 \\0 & 0 & 0 & x_2\end{array}\right)
.\end{equation}
The leading term in the PGB effective Lagrangian is $C{\rm Tr}[D_\mu\Sigma(x)D^\mu\Sigma(x)^\dagger]$. The covariant derivative involves the $SU(3)\times SU(2)_L\times U(1)$ gauge fields and it is determined by fact that $\Sigma(x)$ transforms as a product of fermions. When there are $N_d$ doublets, the constant $C$ and the normalization of $X$ can be obtained by requiring that the expansion of the Lagrangian gives $\frac{1}{2}\partial_\mu\Pi\partial^\mu\Pi+\frac{1}{8}g^2v^2W^0W^0+...$. The result is $C=v^2/(8N_d)$ and 
\begin{equation}
\sum_i x_i^2=\frac{N_d}{2}
\end{equation}
where the sum is over the different colors and flavors. $N_d=3$ for fourth family A while $N_d=4$ for technicolor and fourth family B. We provide a table of the $x_a$'s.
\begin{center}\begin{tabular}{|c|c|c|c|c|}\hline  & $x_{b'}$ & $x_{t'}$ & $x_{\tau'}$ & $x_{\nu'}$ \\\hline 4th family A & 1/2 & 1/2 & 0 & 0 \\\hline 4th family B & $1/\sqrt{3}$ & $1/\sqrt{3}$ & 0 & 0 \\\hline technicolor A & $\sqrt{3}/6$ & $\sqrt{3}/6$ & $-\sqrt{3}/2$ & $-\sqrt{3}/2$ \\\hline technicolor B & $1/\sqrt{6}$ & 0 & $-3/\sqrt{6}$ & 0 \\\hline technicolor C & 0 & $1/\sqrt{6}$ & 0 & $-3/\sqrt{6}$ \\\hline \end{tabular}\end{center}

These $x_a$'s determine the coupling of $\Pi$ to pairs of gauge bosons through the axial-vector anomaly. The resulting width to gluons is
\begin{equation}
\Gamma(\Pi\rightarrow gg)=\frac{m_\Pi^3}{8\pi}\left(\frac{\alpha_s N_{TC}{\cal A}_{gg}}{2\pi v}\right)^2
\label{e1}\end{equation}
where ${\cal A}_{gg}=2x_{t'}+2x_{b'}$. $N_{TC}=1$ for a fourth family and we shall take $N_{TC}=2$ for the technicolor case. In the chiral limit the perturbative corrections to the axial-vector anomaly amplitude vanish and so the corrections are small and of order $(m_\Pi/\mbox{1 TeV})^2$. This differs from the Higgs case where the perturbative corrections are large. The width to photons is
\begin{equation}
\Gamma(\Pi\rightarrow \gamma\gamma)=\frac{m_\Pi^3}{64\pi}\left(\frac{\alpha N_{TC}{\cal A}_{\gamma\gamma}}{2\pi v}\right)^2
\label{e2}\end{equation}
where ${\cal A}_{\gamma\gamma}=\frac{16}{3}x_{t'}+\frac{4}{3}x_{b'}+4x_{\tau'}$. The $gg$ and $\gamma\gamma$ widths of technipions were first obtained in \cite{a2} and for technicolor A our expressions agree with those in \cite{a3} for the original one family technicolor model.

If $\Pi$ has a coupling, $i\lambda_f\Pi\overline{\psi}_f\gamma_5\psi_f$, to a fermion into which it can decay then the width is
\begin{equation}
\Gamma(\Pi\rightarrow \overline{f}f)=N_CK^{QCD}_f\lambda_f^2\frac{m_\Pi}{8\pi}\left(1-\frac{4m_f^2}{m_\Pi^2}\right)^\frac{1}{2}
,\label{e4}\end{equation}
where $N_C=3\;(1)$ for quarks (leptons). $K^{QCD}_f$ is the QCD correction factor ($=1$ for leptons). $\lambda_f$ is related to the origin of the lighter fermion masses (as in extended technicolor \cite{a6}) and various assumptions and conventions can be found in the literature.  In the simplest case the mass of a lighter fermion $f$ is fed down from a single heavy fermion $\psi_a$; then $m_f$ is proportional to the condensate $\langle\overline{\psi}_a\psi_a\rangle$. Since $\Pi$ is a fluctuation of the condensate we can deduce from (\ref{e3}) that $\lambda_f=2x_am_f/v$. An especially simple type of flavor (or extended technicolor) interaction is a `sideways' interaction that simply links fermions of the same type to each other, and so in this case $\lambda_b=2x_{b'}m_b/v$, etc. We shall adopt this as our canonical choice for the $\lambda_f$'s.

More generally the four fermion operators that can contribute to the $b$ mass are $\overline{b'}_Rb'_L\overline{b}_Lb_R$, $\overline{\tau'}_R\tau'_L\overline{b}_Lb_R$,  $\overline{t'}_Lt'_R\overline{b}_Lb_R$ and $\overline{\nu'}_L\nu'_R\overline{b}_Lb_R$ ($+$ h.c.'s). These can all be written in $SU(2)_L\times U(1)$ invariant form with the last two having a different Lorentz structure, and with the last one not contributing if the neutrino condensate is Majorana. Thus if the $b$ mass is proportional to $a_1\langle\overline{b'}b'\rangle+a_2\langle\overline{\tau'}\tau'\rangle+a_3\langle\overline{t'}t'\rangle+a_4\langle\overline{\nu'}\nu'\rangle$ then the $i\Pi\overline{b}\gamma_5b$ coupling is proportional to $x_{b'}a_1\langle\overline{b'}b'\rangle+x_{\tau'}a_2\langle\overline{\tau'}\tau'\rangle-x_{t'}a_3\langle\overline{t'}t'\rangle-x_{\nu'}a_4\langle\overline{\nu'}\nu'\rangle$. Thus if several condensates contribute significantly to a mass then there is no simple relation between the mass and coupling. On the other hand this may lead to flavor changing neutral currents mediated by $\Pi$, and so this situation may be disfavored. The canonical choice is a good starting point.

The values of the constants are as follows.
\begin{center}\begin{tabular}{|c|c|c|c|c|c|}\hline  & $\lambda_b$ & $\lambda_c$ & $\lambda_\tau$ & ${\cal A}_{gg}$ & ${\cal A}_{\gamma\gamma}$ \\\hline 4th family A & $m_b/v$ & $m_c/v$ & 0 & 2 & $10/3$ \\\hline 4th family B & $2m_b/\sqrt{3}v$ & $2m_c/\sqrt{3}v$ & 0 & $4/\sqrt{3}$ & $20/3\sqrt{3}$ \\\hline technicolor A& $m_b/\sqrt{3}v$ &  $m_c/\sqrt{3}v$ & $-\sqrt{3}m_\tau/v$ & $2/\sqrt{3}$ & $-8/3\sqrt{3}$ \\\hline technicolor B& $\sqrt{2}m_b/\sqrt{3}v$ &  0 & $-\sqrt{6}m_\tau/v$ & $\sqrt{2/3}$ & $-16\sqrt{6}/9$ \\\hline technicolor C& 0 &  $\sqrt{2}m_c/\sqrt{3}v$ & 0 & $\sqrt{2/3}$ & $8\sqrt{6}/9$ \\\hline \end{tabular}\end{center}
We see that the $\lambda_f$'s typically differ from standard Higgs Yukawa couplings. For technicolor A we note that our $\lambda_f$'s differ from the couplings used for the `original one family' model in \cite{a3}, which in our notation are $\lambda_f=2m_f/v$, a factor of $2\sqrt{3}$ ($2/\sqrt{3}$) larger than our $\lambda_f$'s for quarks (leptons). These larger couplings are presumably extracted from earlier work where different assumptions are made.  Technicolor B was considered in \cite{a5} and \cite{a4} and our $\lambda_f$'s agree. The $gg$ and $\gamma\gamma$ widths in (\ref{e1}, \ref{e2}) for technicolor B also agree with eq. (35) of \cite{a5}. The `variant one family' model of \cite{a3} is apparently also based on \cite{a5} (the $\lambda_f$'s agree) but then the $gg$ and $\gamma\gamma$ widths used in \cite{a3} are a factor of 4 too small. These differences with \cite{a3} along with our inclusion of the $K^{QCD}_f$ factor in (\ref{e4}) all have the effect of causing our constraints for both technicolor A and B to be stronger than in \cite{a3}.

We can now determine the branching ratios (in \%) and compare to the full Higgs branching ratios \cite{a16}. We use the currently popular mass value $m_\Pi=m_H=125$ GeV. For the $K^{QCD}_f$ factor we use the same factors that enter the Higgs decays, $K^{QCD}_b=.52$,  $K^{QCD}_c=.26$ \cite{a16}.
\begin{center}\begin{tabular}{|p{18ex}|p{6ex}|p{6ex}|p{6ex}|p{6ex}|p{6ex}|}\hline  & $b\overline{b}$ & $c\overline{c}$ &$ \tau^+\tau^-$ & $gg$ & $\gamma\gamma $ \\\hline 4th family A or B & 54 & 2.5 & 0 & 44 & .064 \\\hline technicolor A & 19 & .9 & 19 & 61 & .058  \\\hline technicolor B & 35 & 0 & 36 & 29 & .43 \\\hline technicolor C & 0 & 2.9 & 0 & 97 & .36  \\\hline Higgs & 58 & 2.7 & 6.4 & 8.6 & .23 \\\hline \end{tabular}\end{center}
The production cross section times branching ratios for the processes $gg\rightarrow\Pi\rightarrow\gamma\gamma$ and $gg\rightarrow\Pi\rightarrow\tau^+\tau^-$ can then be obtained in relation to the Higgs. We denote these ratios by $R_\gamma$ and $R_\tau$.
\begin{eqnarray}
R_\gamma&\equiv&R_g\frac{BR(\Pi\rightarrow \gamma\gamma)}{BR(H\rightarrow \gamma\gamma)}\\R_\tau&\equiv&R_g\frac{BR(\Pi\rightarrow \tau^+\tau^-)}{BR(H\rightarrow \tau^+\tau^-)}\\
R_g&\equiv&\frac{\Gamma(\Pi\rightarrow gg)}{\Gamma(H\rightarrow gg)}
\end{eqnarray}
We use the full $\Gamma(H\rightarrow gg)$ from \cite{a16} and obtain
\begin{center}\begin{tabular}{|c|c|c|c|}\hline  & $R_g$ & $R_\gamma$ & $R_\tau$ \\\hline 4th family A & 5.4 & 1.5 & 0 \\\hline 4th family B & 7.2 & 2.0 & 0 \\\hline technicolor A & 7.2 & 1.8 & 22 \\\hline technicolor B & 3.6 & 6.8 & 20 \\\hline technicolor C & 3.6 & 5.7 & 0 \\\hline \end{tabular}\end{center}

The technicolor results show that typically $R_\gamma$ and/or $R_\tau$ are too large given the current sensitivity of the LHC. Relative to a 125 GeV Higgs the current 95\% CL limits are $R_\gamma\lesssim3$ from both ATLAS \cite{a20} and CMS \cite{a19} and $R_\tau\lesssim3.5$ from CMS \cite{a21}. One could argue that the uncertainties in the $\lambda_f$ couplings could weaken the constraints. But one would need significantly larger couplings; the larger $\lambda_f$'s used for technicolor A in \cite{a3} for example give $R_\gamma\approx0.6$ and $R_\tau\approx9$. (These results still differ from \cite{a3} due to the $K^{QCD}_f$'s.) Technicolor C has a better chance, but here one needs to explain why $\Pi$ has the content in (\ref{e5}).

We see that the fourth family case is more viable. We can also consider deviations from the $\lambda_\tau=0$ limit by generalizing the case of fourth family B to $(x_{t'}, x_{b'}, x_{\tau'}, x_{\nu'})=(\sqrt{1/3},\sqrt{1/3},0,0)+(-1/6,1/6,-1/2,1/2)a+(0,0,1/2,1/2)b$. For small $a$ and $b$ we find $R_\gamma\approx2.0-3.4a+2.1b$ and $R_\tau\approx5.2(a-b)^2$.

We have omitted the top quark contribution to the triangle graph coupling $\Pi$ to $gg$ and $\gamma\gamma$. The $\Pi$ coupling to the top quark is model dependent and could result in constructive or destructive interference with the other contributions to the $gg$ and $\gamma\gamma$ couplings \cite{a3}. An additional complication for the fourth family is that $\Pi$ may itself contain a top quark component which thus produces a direct coupling to the top quark. There is still the coupling induced by a 4-fermion operator such as $\overline{b'}_Lb'_R\overline{t}_Lt_R$.\footnote{Note that we have chosen an operator which is not generated by a simple `sideways' interaction. This helps to alleviate \cite{a12} problems \cite{a13} associated with trying to use sideways interactions to generate a large enough top mass.} A plausible situation, given the $\Pi$ mass, is for a modified $\Pi$ generator $X$ to correspond to a symmetry of the 4-fermion operator responsible for the top quark mass \cite{a11}. This can result in a cancellation between the two contributions to the $i\Pi\overline{t}\gamma_5t$ coupling. In any case the top quark contribution is uncertain and it could end up increasing or decreasing $R_\gamma$.

The axial-vector anomaly amplitude contains a three gluon piece which leads to some NLO corrections to the above results. The process $\Pi\rightarrow ggg$ adds a decay mode and thus reduces $BR(\Pi\rightarrow \gamma\gamma)$. On the other hand the process $gg\rightarrow\Pi g$ increases the $\Pi$ production rate, and so these effects partially cancel in $R_\gamma$. Finally we note that the higher order processes that yield the production of $\Pi+2$ jets are also of interest, since this production could be mistaken for weak boson fusion production of the Higgs. There is no tree level weak boson fusion production of the $\Pi$.

We conclude that a fourth family description of a 125 GeV pseudoscalar state is presently viable (see also \cite{a18,a18a}), while a technicolor description is more difficult. Stepping back we note that there are many possible experimental signatures of theories of new strong interactions at the TeV scale, and they have been extensively explored for more than three decades. From the viewpoint of that literature it is quite unexpected to find that the first hint of such theories could occur via a neutral color singlet state decaying into two photons. This is made possible by the intensely focused search for the Higgs.

\section*{Acknowledgments}
This work was supported in part by the Natural Science and Engineering Research Council of Canada.

\end{document}